\documentclass[twocolumn,aps,prl,groupedaddress]{revtex4-1}
\usepackage[utf8]{inputenc}
\usepackage{color}
\usepackage{array}
\usepackage{mathtools}
\usepackage{amsmath}
\usepackage{graphicx}
\usepackage{subscript}

\makeatletter

\providecommand{\tabularnewline}{\\}

\usepackage[euler]{textgreek}
\usepackage[caption=false]{subfig}
\usepackage{hhline}
\usepackage[bottom]{footmisc}
\@ifundefined{showcaptionsetup}{}{%
 \PassOptionsToPackage{caption=false}{subfig}}
\usepackage{subfig}
\makeatother

\begin{document}

\title{A high-performance optical lattice clock based on bosonic atoms}

\author{S. Origlia, M.S. Pramod, S. Schiller}

\affiliation{Institut für Experimentalphysik, Heinrich-Heine-Universität Düsseldorf,
40225 Düsseldorf, Germany}

\author{Y. Singh, K. Bongs}

\affiliation{University of Birmingham, Birmingham B15 2TT, United Kingdom}

\author{R. Schwarz, A. Al-Masoudi, S. Dörscher, S. Herbers, S. Häfner, U.
Sterr and Ch. Lisdat}

\affiliation{Physikalisch-Technische Bundesanstalt, 38116 Braunschweig, Germany}

\date{February 24, 2018}
\begin{abstract}
Optical lattice clocks with uncertainty and instability in the $10^{-17}$-range
and below have so far been demonstrated exclusively using fermions.
Here, we demonstrate a bosonic optical lattice clock with $3\times10^{-18}$
instability and $2.0\times10^{-17}$ accuracy, both values improving
on previous work by a factor 30. This was enabled by probing the clock
transition with an ultra-long interrogation time of $4$~s, using
the long coherence time provided by a cryogenic silicon resonator,
by careful stabilization of relevant operating parameters, and by
operating at low atom density.   This work demonstrates that bosonic
clocks, in combination with highly coherent interrogation lasers,
are suitable for high-accuracy applications with particular requirements,
such as high reliability, transportability, operation in space, or
suitability for particular fundamental physics topics. As an example,
we determine the \textsuperscript{88}Sr – \textsuperscript{87}Sr
isotope shift with 12~mHz uncertainty.
\end{abstract}
\maketitle

\section{Introduction}

\vskip 13pt

Lattice optical clocks~\citep{Katori2003} have made strong progress
in the past decade, both in terms of accuracy and stability. Atomic
species intensely investigated so far are strontium, ytterbium, mercury,
and magnesium. Best performances today include instability at low-$10^{-16}$
level at $\tau=1\,$s integration time~\citep{Nicholson2012,Hinkley2013,AlMasoudi2015},
with the lowest value of $6\times10^{-17}/\sqrt{\tau/\mathrm{s}}$
achieved with dead-time-free interleaved interrogation of two atomic
ensembles~\citep{Schioppo2017}. The lowest uncertainty has been
estimated for an individual Sr clock, $2.1\times10^{-18}$~\citep{Nicholson2015},
while direct lattice clock comparisons have achieved agreement with
a fractional uncertainty of $4.4\times10^{-18}$ at best~\citep{Ushijima2015}.
Lattice clocks operated in different metrological institutes have
been compared over long-distance links~\citep{Calonico2014,Takano2016,Lisdat2016,Predehl2012}
and the feasibility of using lattice clocks for the realization of
time scales has been demonstrated~\citep{Lodewyck2016,Grebing2016}.
First applications have also been reported: a transportable Sr~lattice
clock~\citep{Koller2017} has successfully completed measurement
campaigns away from its home base~\citep{Grotti2018}. Relativistic
geodesy with uncertainty at the 5~cm-level has been implemented by
comparing two clocks in different laboratories at 15~km distance~\citep{Takano2016}.
A strontium lattice clock is also foreseen for the ``Space Optical
Clock on the ISS'' mission of ESA~\citep{Schiller2012a}.

\vskip 4pt

Lattice clocks can be operated with bosonic or fermionic isotopes.
Bosons exhibit some disadvantages compared with fermions, which has
significantly slowed down their use for metrological applications:
(i) the presence of \textit{s}-wave collisions~\citep{Lisdat2009},
which for spin-polarized fermions are suppressed, and which cause
a significant frequency shift if uncontrolled, and (ii) a completely
forbidden transition between the \textsuperscript{1}S\textsubscript{0}
and the \textsuperscript{3}P\textsubscript{0} clock states, which
requires the application of one or more additional external fields
to enable driving the transition~\citep{Taichenachev2006,Hong2005}.
These fields also cause a significant frequency shift~\citep{Hong2005,Taichenachev2006,Baillard2007,Takano2017}.
On the other hand, the lifetime of the upper level is significantly
shorter in fermions than in the bosonic isotopes, a fact that will
become a limitation once the next generation of ultrastable lasers~\citep{Matei2017}
is employed. 

\vskip 4pt

Turning to the specific case of strontium, the isotopes used so far
are \textsuperscript{87}Sr (fermion) and \textsuperscript{88}Sr
(boson). The bosonic isotope is attractive for realizing a simplified
lattice clock, e.g. for transportation or for use in space on a satellite,
where robustness and reliability are essential. Compared with \textsuperscript{87}Sr,
the atom cooling and clock spectroscopy are conceptually and technically
simpler, thanks to the higher natural isotopic abundance (83\% for
\textsuperscript{88}Sr versus 7\% for \textsuperscript{87}Sr) and
absence of hyperfine structure~\citep{Katori2003,Mukaiyama2003}.
Furthermore, the Stark shift cancellation wavelength (``magic''
lattice wavelength) in the bosonic isotope is basically insensitive
to changes of the external magnetic field or of the lattice polarization
axis~\citep{Westergaard2011}. It is also important to develop Sr
bosonic clocks further in view of the fact that isotope shifts of
all bosonic Sr atoms (\textsuperscript{84}Sr, \textsuperscript{86}Sr,
\textsuperscript{88}Sr, \textsuperscript{90}Sr) are of significant
interest for the search of physics beyond the standard model~\citep{Delaunay2016}.
In this context, a highly accurate measurement of the \textsuperscript{88}Sr–\textsuperscript{87}Sr
isotope shift has been recently performed in a dedicated setup, sharing
large common-mode perturbations~\citep{Takano2017}.

\vskip 4pt

Although the potential of bosonic lattice clocks had been foreseen
a long time ago~\citep{Taichenachev2006}, no experimental proof
could be given, so far. In this paper, we demonstrate that a high-performance
lattice clock is indeed feasible with a bosonic isotope.

\section{Magnetically induced spectroscopy}

The optical excitation of the \textsuperscript{1}S\textsubscript{0}-\textsuperscript{3}P\textsubscript{0}
clock transition in a bosonic atom can be enabled applying a bias
magnetic field $B$. This results in a Rabi frequency for the interrogation
of the clock transition given by $\Omega_{\mathrm{R}}/2\pi=\alpha\sqrt{I}\left|B\right|$,
with the coupling coefficient \textit{$\alpha=198\,\text{\text{\textrm{\ensuremath{\mathrm{{Hz/T\sqrt{mW/cm^{2}}}}}}}}$}
for \textsuperscript{88}Sr~\citep{Taichenachev2006} and the clock
interrogation wave intensity~$I$. For typical $B$-fields, $I$
is 2 to 3 orders of magnitude stronger compared to fermionic lattice
clocks~\citep{Taichenachev2006}. The intensity leads to a probe
light shift $\Delta\nu_{\mathrm{L}}=k\,I$, with $k=-18\mathrm{\text{ }mHz/(mW/cm^{2})}$~\citep{Taichenachev2006},
while the magnetic field leads to a 2\textsuperscript{nd}-order Zeeman
shift $\Delta\nu_{\mathrm{B}}=\beta\,B^{2}$, with $\beta=-23.8(3)\mathrm{\text{ }MHz/T^{2}}$~\citep{Taichenachev2006}.\textcolor{black}{{}
Thus the Rabi frequency is proportional to }$\sqrt{\Delta\nu_{\mathrm{L}}\cdot\Delta\nu_{\mathrm{B}}}$\textcolor{black}{.}
The clock transition interrogation time, in case of Rabi spectroscopy,
can be expressed as $T_{\pi}=\pi/\Omega_{R}$: therefore, the Rabi
frequency, and consequently $\Delta\nu_{L}$ and $\Delta\nu_{B}$,
can be reduced using a clock laser that supports longer $T_{\pi}$.\textcolor{black}{{}
The two shifts and the interrogation time are related by $\left|\varDelta\nu_{\mathrm{L}}\varDelta\nu_{\mathrm{B}}\right|=2.8\,T_{\pi}^{-2}$.
By choosing $B$ and $I$ appropriately, the minimum realizable shift
magnitude is }%
\mbox{%
$\lvert\Delta\nu_{\mathrm{L}}+\Delta\nu_{\mathrm{B}}\rvert_{\mathrm{min}}=3.3/T_{\pi}$%
}\textcolor{black}{.} Furthermore, a long interrogation time reduces
the effect of detection noise (including quantum projection noise)
on the clock instability since these contributions scale as $1/\sqrt{T_{\pi}}$.
As a consequence, operating the clock with low atom number with moderate
degradation of the stability becomes possible.
\begin{figure}
\begin{centering}
\includegraphics[scale=0.48]{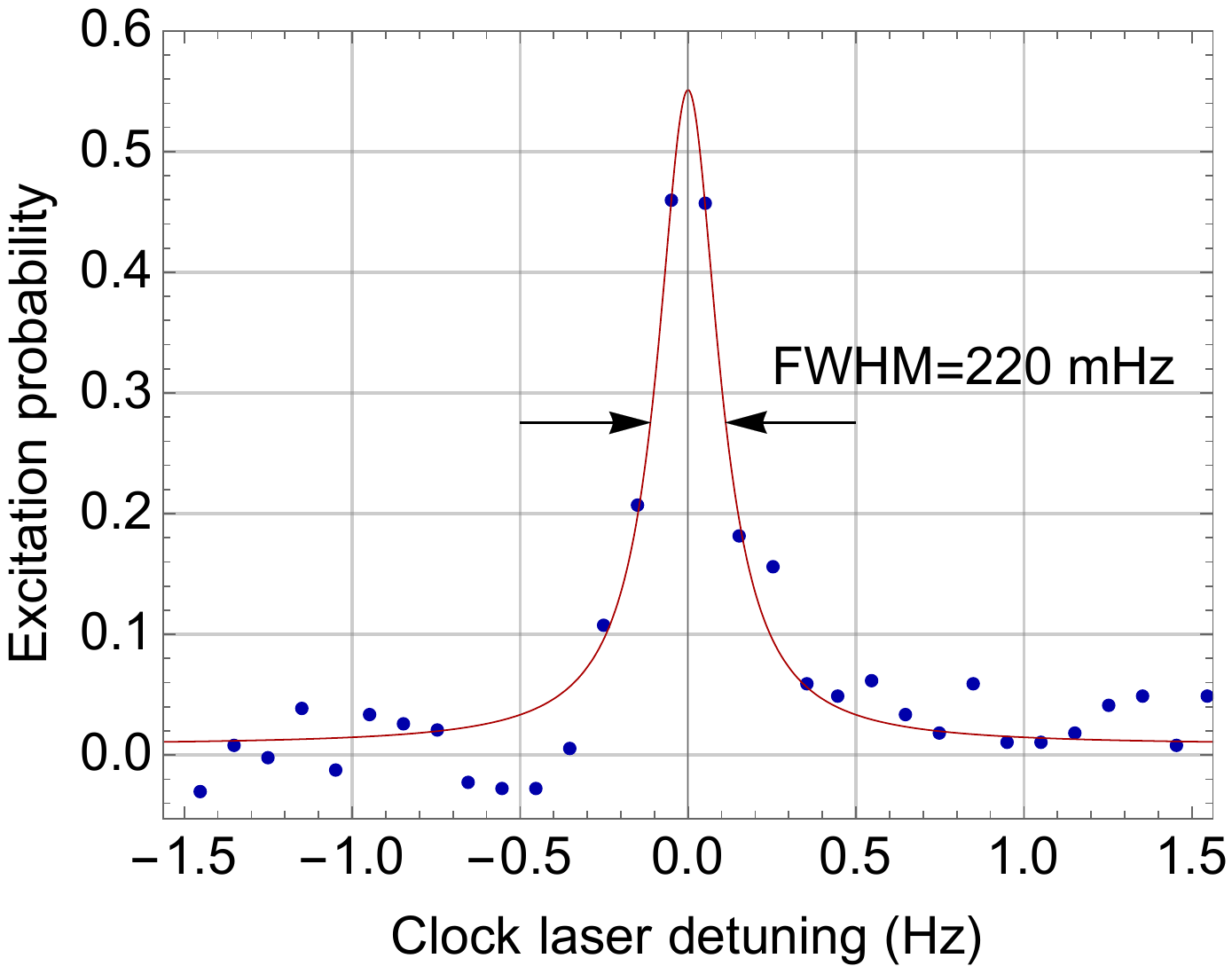} 
\par\end{centering}
\caption{Typical \protect\protect\protect\textsuperscript{88}Sr clock transition
line for $T_{\pi}=4.0$~s, bias field $B_{\mathrm{exp,0}}\approx0.21$~mT,
clock laser intensity $I_{\mathrm{exp,0}}\approx28$~mW/cm\protect\protect\protect\textsuperscript{2}.
The line is a single scan (i.e., no averaging), with a total scan
time of 165~s. The red line represents a Lorentzian fit with 220~mHz
full width at half maximum (FWHM) linewidth. \label{fig:clockTransition-1}}
\end{figure}

\section{The experimental apparatus}

The lattice clock apparatus with its cooling and manipulation lasers
is described in~\citep{Bongs2015,Origlia2016}. Atoms are cooled
and trapped in a 1D vertically oriented optical lattice (magic wavelength:
813 nm, %
\mbox{%
$\sim40\:\mu$%
}m waist radius). The lifetime of the atoms in the lattice is \textgreater{}5~s.
The clock laser (698~nm) is pre-stabilized on a 10~cm long transportable
cavity~\citep{Vogt2011a} and phase-locked to a stationary clock
laser stabilized on a 48~cm long reference cavity~\citep{Haefner2015}.
For most of the measurements, the latter was phase-locked, using a
transfer-lock scheme~\citep{Stenger2002}, to a 1540~nm laser locked
to a cryogenic silicon resonator~\citep{Matei2017}, which exhibits
less than 10~mHz linewidth and $4\times10^{-17}$ instability at
1~s integration time. The clock laser radiation is delivered to the
atoms via a phase-noise-cancelled optical fiber~\citep{Falke2012}.
The clock laser waist radius is approximately $105~\mu$m and provides
a fairly homogeneous intensity profile across the atomic sample.

\section{Spectroscopy and clock instability}

We investigated atom interrogation times up to $T_{\mathrm{\mathit{\pi}}}=8$~s.
\textit{$T_{\mathrm{\mathit{\pi}}}=4~$}s ($\Omega_{\mathrm{\mathrm{R}}}/2\pi=0.125~$Hz)
was chosen as optimum value, the longest for which reliable Fourier-limited
clock transition linewidths were observed (Fig.~\ref{fig:clockTransition-1}).
This leads to a total cycle time of 5.3~to~6.3~s depending on the
operating conditions. The observed linewidth of 0.22~Hz is in agreement
with the theoretically expected value $0.8/T_{\mathrm{\mathit{\pi}}}\simeq0.2$~Hz~\citep{Dick1988}.
The reason for the limited contrast ($\sim$60\%) can be attributed
to collisional effects (see Section~\ref{subsec:Cold-collisional-shift}).

In order to control the probe light shift and Zeeman shift below the
$1\times10^{-17}$ (4.3~mHz) level, the clock laser beam power and
the current $I_{\mathrm{B,0}}$ in the bias field coils are actively
stabilized. For the clock laser power we used a combination of an
analog and digital power stabilization, acting on the RF power feeding
an acousto-optic frequency shifter (AOM) in the clock laser breadboard.
The digital stabilization serves as integrator of the analog error
signal spanning several experimental cycles. It minimizes lock errors
of the analog servo when the beam is turned on. With this system we
can achieve a long-term beam power fractional instability below $1\times10^{-3}$
over a few days, corresponding to $<2\times10^{-18}$ for the fractional
shift, for our interrogation parameters (see below). The current stabilization
is based on a digital multimeter (DMM) measuring the current; the
DMM reading is fed into a digital PID control, which steers the external
control voltage of the power supply once every clock cycle. Based
on the DMM's specifications, the expected fractional instability of
the 2\textsuperscript{nd}-order Zeeman shift, using $T_{\mathit{\pi}}=4$~s
and $I_{\mathrm{\mathrm{\mathrm{\mathit{\mathrm{B}},0}}}}=+215$~mA,
is below $2\times10^{-4}$ (over 24~hours), corresponding to a $3.1\times10^{-18}$
uncertainty on the frequency shift.

Any practical combination of clock laser intensity and bias field
strength matching the desired Rabi frequency can be used. Based on
the performance of the clock laser power and bias field current stabilizations,
we chose a clock laser intensity $I_{\mathrm{exp,0}}\approx28$~mW/cm\textsuperscript{2}
and a bias field $B_{\mathrm{exp,0}}\approx0.21$~mT, leading to
shifts $\Delta\nu_{\mathrm{L,0}}=k\,I_{\mathrm{exp,0}}\approx-0.50$~Hz
($5.8\times10^{-16}$ in fractional units) and $\Delta\nu_{\mathrm{\mathrm{\mathit{\mathit{B}},0}}}=\beta\,B_{\mathrm{exp,0}}^{2}\approx-1.03$~Hz
($2.4\times10^{-15}$).

\vskip 4pt

\begin{figure}
\begin{centering}
\includegraphics[scale=0.41]{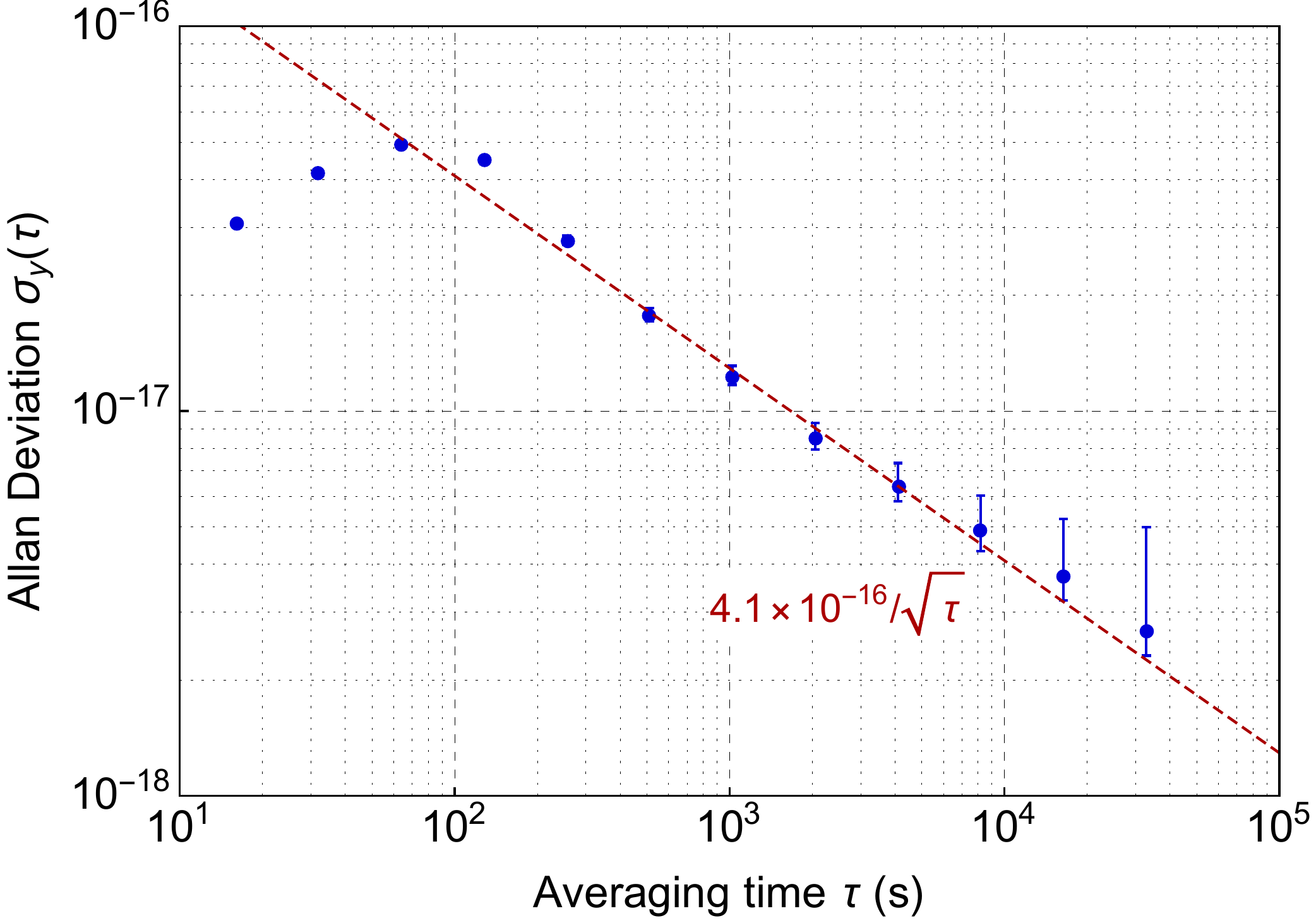} 
\par\end{centering}
\centering{}\caption{Allan deviation of the fractional frequency offset between the \protect\protect\textsuperscript{88}Sr
clock and the PTB \protect\protect\textsuperscript{87}Sr clock \label{fig:longInstabilityWithStat} }
\end{figure}

In order to determine the clock instability, the bosonic clock was
compared to the fermionic \textsuperscript{87}Sr clock at PTB~\citep{AlMasoudi2015,Grebing2016}.
Although the clock lasers of both clocks are prestabilized to the
same cryogenic silicon cavity, on long time scales ($\tau>200$~s),
they are steered to the respective atomic references and, thus, independent.
The combined instability is $4.1\times10^{-16}/\sqrt{\tau/\mathrm{s}}$,
and averages down to the $3\times10^{-18}$ level~(Fig.~\ref{fig:longInstabilityWithStat}).
A number of measurements were acquired over the time span of a few
months with similar results.

\section{Systematic shift evaluation}

Table~\ref{Tab:UncBudget} summarizes the uncertainty budgets of
the bosonic and fermionic clock~\citep{Falke2014}. The systematic
shifts of the bosonic clock were controlled and determined in the
following way.
\begin{table}
\begin{centering}
\begin{tabular}{l>{\centering}p{9mm}>{\centering}p{9mm}>{\centering}p{9mm}>{\centering}p{9mm}}
\noalign{\vskip0.4cm}
\hhline{=====} & \multicolumn{2}{c}{\textsubscript{}\textsuperscript{88}Sr clock} & \multicolumn{2}{c}{\textsuperscript{87}Sr clock}\tabularnewline
Effect & $\Delta\nu$  & $u$  & $\Delta\nu$  & $u$\tabularnewline
\hline 
 BBR shift  & 523.2  & 0.8  & 492.2  & 1.5\tabularnewline
BBR oven  & 0  & 0  & 0.9  & 0.9\tabularnewline
Lattice shifts  & 2.2  & 1.1  & 0.9  & 0.4\tabularnewline
Probe light shift ($\Delta\nu_{\mathrm{L}}$)  & 96.1  & 1.3  & 0.0  & 0.0\tabularnewline
Cold collisions ($\Delta\nu_{\mathrm{LP}}$)  & 0.6  & 0.3  & 0.0  & 0.2\tabularnewline
2\textsuperscript{nd}-order Zeeman shift ($\Delta\nu_{\mathrm{B}}$)  & 209.7  & 0.5  & 3.4  & 0.1\tabularnewline
Tunneling  & 0  & 0  & 0.0  & 0.3\tabularnewline
Background gas collisions  & 0.13  & 0.13  & 0.8  & 0.8\tabularnewline
DC-Stark shift  & 0.2  & 0.2  & 0.2  & 0.1\tabularnewline
\noalign{\vskip0.2cm}
Total  & 827.1  & 2.0  & 498.4  & 2.0\tabularnewline
\hhline{=====} \\[-2.5em] &  &  &  & \tabularnewline
\end{tabular}
\par\end{centering}
\caption{Uncertainty budget for the \protect\protect\protect\textsuperscript{88}Sr
and \protect\protect\protect\textsuperscript{87}Sr clocks. All
numbers are expressed in parts per 10\protect\protect\textsuperscript{17}.}
\label{Tab:UncBudget} 
\end{table}

\subsection{Blackbody radiation shift}

For the evaluation of the blackbody radiation (BBR) shift, the temperature
of the chamber is monitored by 17 temperature sensors (10 of them
on the chamber windows). Thanks to the small size of the chamber (outer
size 50~$\times$~50~$\times$~20~mm\textsuperscript{3}) and
of the MOT coils, a passive cooling system based on heat pipes is
sufficient for dissipating most of the heat (about 8~W) produced
by the MOT coils. We do not use any active temperature stabilization,
but rely on the high stability of the laboratory temperature. The
resulting difference between the warmest ($T_{\mathrm{max}}$) and
coldest point ($T_{\mathrm{min}}$) of the vacuum chamber is between
250 and 400~mK, depending on operational parameters. Assuming a uniform
probability distribution for the temperature experienced by the atoms,
the mean temperature is $T_{\mathrm{\mathrm{avg}}}=(T_{\mathrm{max}}+T_{\mathrm{min}})/2$
with uncertainty $(T_{\mathrm{ma}x}-T_{\mathrm{min}})/\sqrt{12}$~\citep{BIPM2008}.
The BBR shift is computed from~\citep{Middelmann2012a,Nicholson2015}.
An atomic beam shutter upstream from the atom chamber is closed during
the atom interrogation cycle and shields the atoms from oven BBR.

\subsection{Lattice light shift}

For the lattice light shift evaluation we used the expression given
in~\citet{Katori2015}:

\[
\Delta\nu_{\mathrm{l}}=\zeta\left(n+\frac{1}{2}\right)\left(\frac{U_{\mathrm{0}}}{E_{\mathrm{r}}}\right)^{1/2}+
\]

\[
-\left(\frac{\partial\Delta\alpha^{\mathrm{E1}}}{\partial\nu}\Delta\nu_{\mathrm{lat,m}}+\frac{3}{4}\Delta k_{\mathrm{H}}(2n^{2}+2n+1)\right)\frac{U_{\mathrm{0}}}{E_{\mathrm{r}}}+
\]

\begin{equation}
+\Delta k_{\mathrm{H}}(2n+1)\left(\frac{U_{\mathrm{0}}}{E_{\mathrm{r}}}\right)^{3/2}-\Delta k_{\mathrm{H}}\left(\frac{U_{\mathrm{0}}}{E_{\mathrm{r}}}\right)^{2},\label{eq:latticeShift}
\end{equation}

where $\zeta=(\frac{\partial\Delta\alpha^{\mathrm{E1}}}{\partial\nu}\Delta\nu_{\mathrm{lat,m}}-\Delta\alpha^{\mathrm{E2,M1}})$,
with $\Delta\alpha^{\mathrm{E1}}=\alpha_{\mathrm{e}}^{\mathrm{E1}}-\alpha_{\mathrm{g}}^{\mathrm{E1}}$
differential dipole polarizability, $\nu_{\mathrm{lat}}$ lattice
frequency, $\Delta\nu_{\mathrm{lat,m}}$ detuning of the lattice frequency
from the magic value, and $\Delta\alpha^{\mathrm{E2,M1}}=(\alpha_{\mathrm{e}}^{\mathrm{E2}}+\alpha_{\mathrm{e}}^{\mathrm{M1}})-(\alpha_{\mathrm{g}}^{\mathrm{E2}}+\alpha_{\mathrm{g}}^{\mathrm{M1}})=0.0(3)$~mHz~\citep{Westergaard2011}
differential multipolar polarizability. The partial derivative of
the differential dipole polarizability is given by $\frac{\partial\Delta\alpha^{\mathrm{E1}}}{\partial\nu_{\mathrm{lat}}}=19.3\times10^{-12}$~(Hz~Hz\textsuperscript{-1})~\citep{Katori2015},
$n$ is the average motional state occupation, $U_{0}$ is the trap
depth, and $E_{\mathrm{r}}=\hbar k_{\mathrm{lat}}^{2}/(2\pi)$ is
the recoil energy, with $k_{\mathrm{lat}}$ the lattice wavenumber.
$\Delta k_{\mathrm{H}}$ is the hyperpolarizability coefficient, whose
most accurate published value is 0.45(10)~$\mu$Hz$/E_{\mathrm{r}}^{2}$~\citep{LeTargat2013}.
The bosonic isotope is not affected by tensor or vector shifts, due
to the absence of a hyperfine structure.

\vskip 1pt

The lattice light shift is measured by an interleaved clock operation
at two different lattice depths, 100~$E_{\mathrm{r}}$ (shallow lattice)
and 157~$E_{\mathrm{r}}$ (deep lattice), $E_{\mathrm{r}}$ being
the lattice photon recoil energy. The difference in the transition
frequency is thereby measured with a fractional uncertainty of $6.5\times10^{-18}$,
after $\sim10000$~s of averaging. This provides the detuning from
the magic wavelength, solving Equation~\eqref{eq:latticeShift} for
$\Delta\nu_{\mathrm{lat,m}}$, which turned out to be 11~MHz. Sideband
spectroscopy was used to evaluate the parameter $n$ (1.4) and the
lattice depth $U_{0}$ (129~$E_{\mathrm{r}}$) under operating conditions.
All the other parameters are known. The lattice shift is than evaluated
for $U_{0}$ and is reported in Tab.~\ref{Tab:UncBudget}.

\subsection{2\protect\protect\textsuperscript{nd}-order Zeeman shift}

\begin{figure}
\begin{centering}
\includegraphics[scale=0.46]{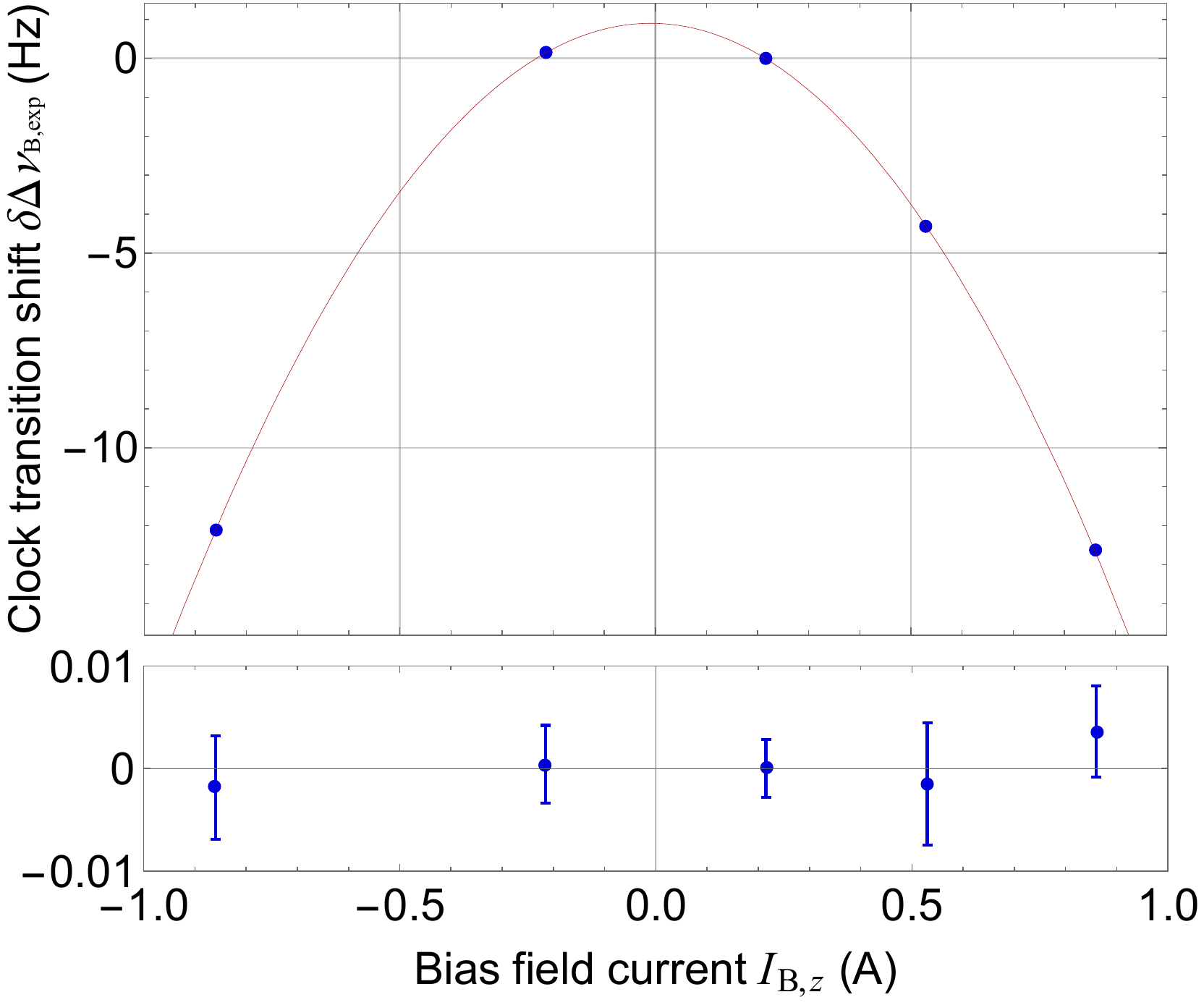} 
\par\end{centering}
\centering{}\caption{Measurement of the 2\protect\protect\protect\textsuperscript{nd}-order
Zeeman shift, and fit residuals, relative to the frequency shift at
the operating point $I_{\mathrm{\mathit{\mathrm{B}},0}}=+215$~mA,
which is $\Delta\nu_{\mathrm{\mathit{\mathrm{B}},exp,0}}=898.9(29)$~mHz.
\label{fig:zeemanShiftMeas-1}}
\end{figure}

Fig.~\ref{fig:zeemanShiftMeas-1} shows the measurement of the 2\textsuperscript{nd}-order
Zeeman shift induced by the coupling field $B$ (parallel to the lattice
polarization axis, $z$-axis). Each point is acquired from an interleaved
measurement, with two different currents $I_{\mathrm{ref},z}$ (reference
current) and $I_{\mathit{\mathrm{B}},z}$. A system of switches allows
inverting the current in the coils. The interrogation time $T_{\pi}$
is set to match the different Rabi frequencies of the cycles. The
data are fit with the function $\delta\Delta\nu_{\mathrm{\mathit{\mathrm{B}},exp}}=\gamma((I_{\mathrm{ref},z}-I_{0,z})^{2}-(I_{\mathrm{B},z}-I_{0,z})^{2})$,
where $I_{0,z}$ is the current necessary in order to compensate for
the external offset field in $z$ direction (maximum of the parabola).
This gives $I_{0,z}=-0.009(2)$~A and $\gamma=17.94(1)$~Hz/A\textsuperscript{2}:
thus the shift at the reference current is $\Delta\nu_{\mathrm{\mathit{\mathrm{B}},exp,0}}=\gamma(I_{0,z}-I_{\mathrm{ref},z})^{2}=-900.2(22)$~mHz
(see Tab.~\ref{Tab:UncBudget}).

Before measuring the shift induced by the field in the $z$-axis,
we compensate for the offset fields in the $x$ and $y$-axes. This
is done by repeating the 2\textsuperscript{nd} order Zeeman shift
measurement, using two pairs of compensation coils with axes perpendicular
to $B$, and finally setting their currents at the values which minimized
the shift. It is important that the compensation of the perpendicular
magnetic fields is done before the measurement of the shift induced
by the coupling field~$B$: in fact, in case the field produced by
the compensation coils at the atom position is not perfectly perpendicular
to $B$, a change of the current in the $x$ and $y$ compensation
coils would result in a change of the offset field in the $z$ direction
(and, consequently, of the parameter $I_{0,z}$ in the fit function).

\begin{figure}
\begin{centering}
\includegraphics[scale=0.66]{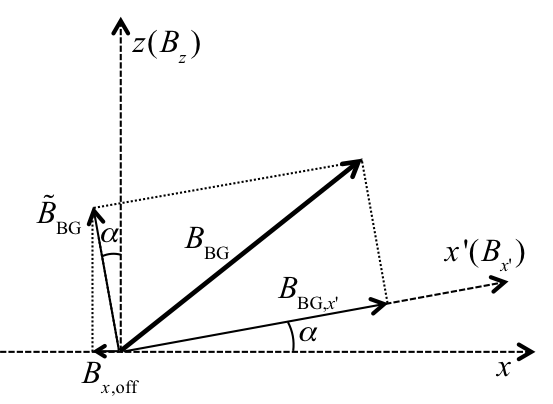} 
\par\end{centering}
\centering{}\caption{Schematic of the magnetic field components for the 2\protect\protect\textsuperscript{nd}-order
Zeeman shift measurement. Details are given in the text. \label{fig:zeemanShiftMeas-1-1}}
\end{figure}

The same circumstance introduces an additional issue which is depicted
in Fig.~\ref{fig:zeemanShiftMeas-1-1}. 
$B_{\mathrm{BG}}$ is the background offset field, and $x'$ is the
direction of the field $B_{\mathrm{\mathit{x}}'}$ produced by the
$x$-compensation coils. $x'$ is misaligned by an angle $\alpha$
from the $x$-direction. By minimizing the 2\textsuperscript{nd}-order
Zeeman shift as a function of the field produced by $x$-compensation
coils, only the component $B_{\mathrm{BG,\mathit{x}'}}$ of the background
field is nulled. The remaining component $B_{\mathrm{BG,\mathit{z}'}}$
introduces a residual offset field $B_{\mathrm{off,\mathit{x}}}$
along the $x$-axis, which results in a residual shift. In order to
estimate $B_{\mathrm{off,\mathit{x}}}$, we evaluate the angle $\alpha$.
For this purpose we determine $B_{\mathrm{off,\mathit{z}}}$ as function
of $B_{x'}$. This is done by measuring the quadratic Zeeman shift
for two opposite values of $B$ (bias field in $z$-direction) and
for $B_{x'}=0$ and $B_{x'}=B_{\mathrm{\mathit{x}',max}}$. We obtain
$B_{\mathrm{off,\mathit{x}}}=B_{\mathrm{off,}z}/\tan\alpha$ and,
from that, the residual shift. The same is repeated for the $y$-direction.
In both cases, the residual shift is much smaller than $1\times10^{-18}$.
If these remaining shifts were larger, they could be further reduced
by iteration of the compensation procedure.

\subsection{Probe light shift}

\begin{figure}
\begin{centering}
\includegraphics[scale=0.48]{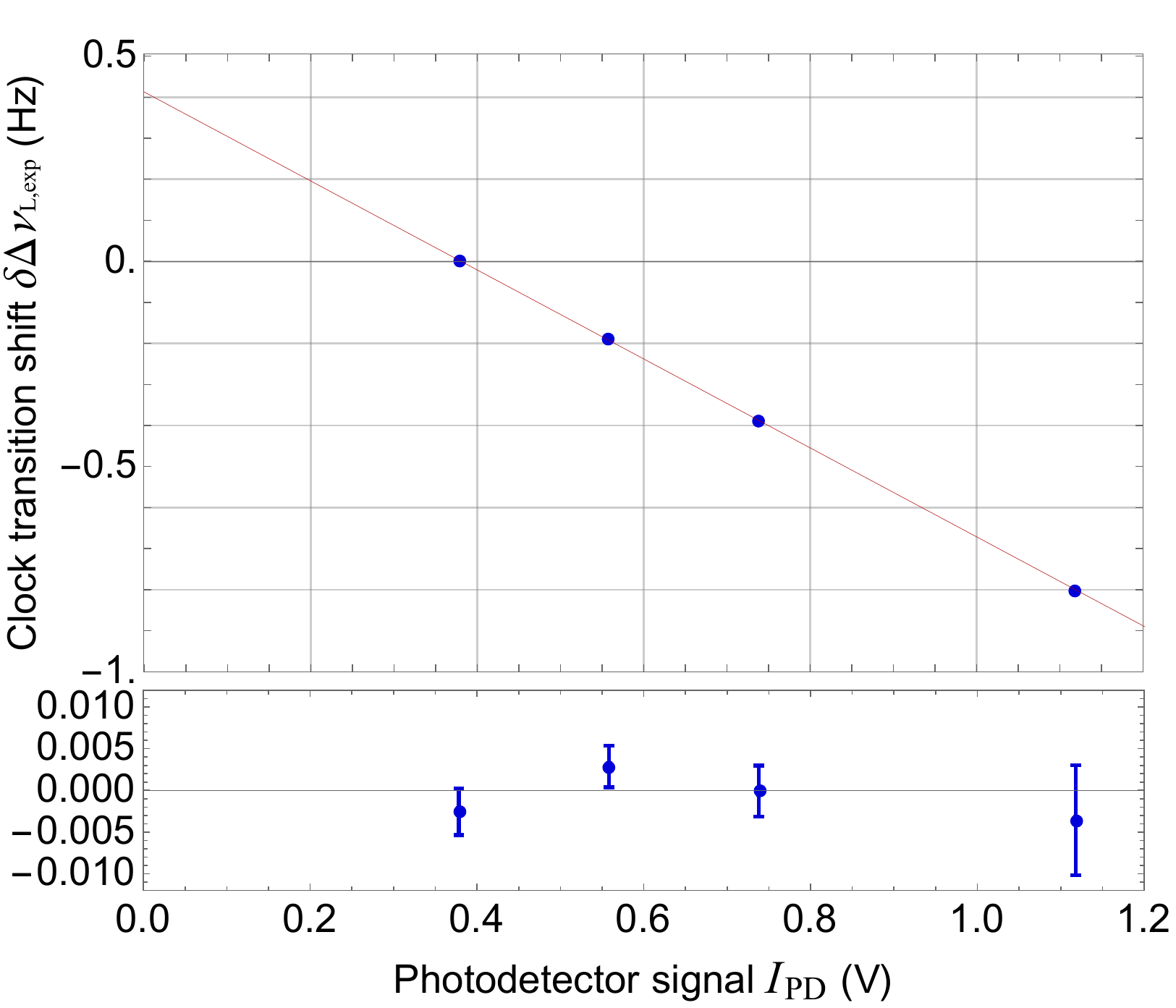} 
\par\end{centering}
\centering{}\caption{Measurement of the probe light shift and fit residuals, relative to
the shift at the operating point $I_{\mathrm{PD,0}}=378$~mV. \label{fig:probeLightShiftMeas-1-1}}
\end{figure}

Similarly, the probe light shift is measured (Fig.~\ref{fig:probeLightShiftMeas-1-1}).
 The $\pi$-pulse condition is again fulfilled, for changing Rabi
frequency due to intensity variation, by adapting $T_{\pi}$. The
data are fit with a linear function. At the operating value, the measurement
yields a shift $\Delta\nu_{\mathrm{L,exp,0}}=-413.0$(53)~mHz. The
difference compared with the expected shift of $-0.50$~Hz is attributed
to the uncertainty of the intensity stabilization photodetector's
calibration, and to non-perfect overlap between the clock and the
lattice beam.
\begin{figure}
\begin{centering}
\includegraphics[scale=0.43]{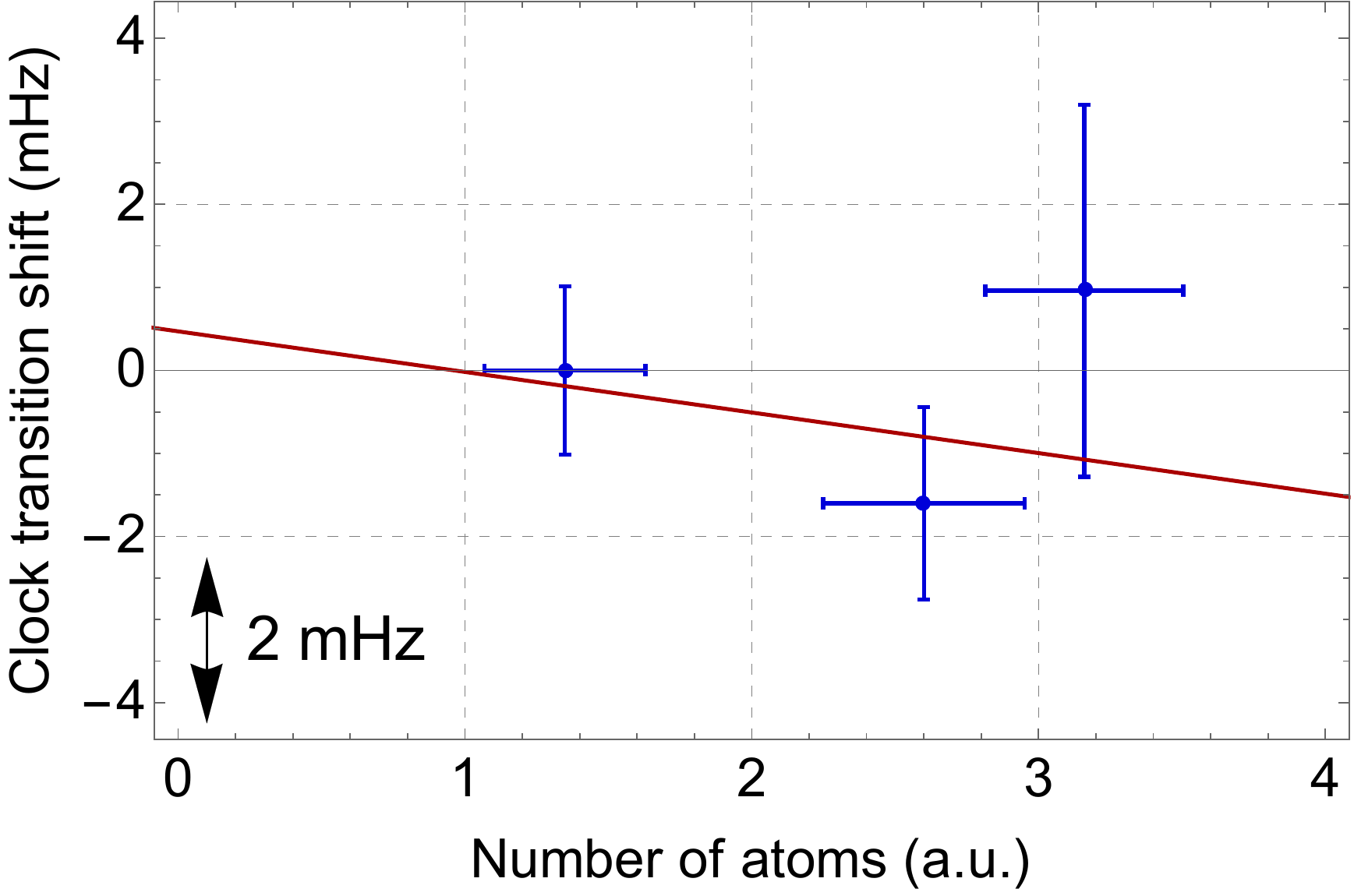} 
\par\end{centering}
\centering{}\caption{Determination of the collisional shift. The standard deviation is
used for the uncertainty on the atom number.\label{fig:collisionalShift}}
\end{figure}

\subsection{Cold collisional shift\label{subsec:Cold-collisional-shift}}

\subsubsection{``Standard'' determination}

For the evaluation of the density (cold collision) shift, we performed
at first a ``standard'' measurement, varying the number of atoms
trapped in the lattice by changing the Zeeman slower beam power. Due
to the line broadening occurring at higher atomic densities (see next
section), it is not possible to significantly increase the atom number
compared to the operating value. This limits the range of data available
for the shift determination. The measurement is performed by running
the clock with a single atomic servo (i.e., not in interleaved mode)
and using the PTB Sr clock as flywheel. Fig.~\ref{fig:collisionalShift}
shows the result. During the measurement at the lowest atom number
(left-most data point) the number of atoms is actively stabilized,
by correcting appropriately the 461~nm slower wave power. From a
fit to the data the cold collisional shift for the atom number under
operating conditions is $0.5(22)$~mHz ($1(5)\times10^{-18}$).

\subsubsection{Alternative evaluation: lineshape analysis\label{subsec:lineshape}}

\begin{figure}
\noindent \begin{centering}
\noindent\begin{minipage}[t]{1\columnwidth}%
\subfloat[]{\begin{centering}
\includegraphics[scale=0.37]{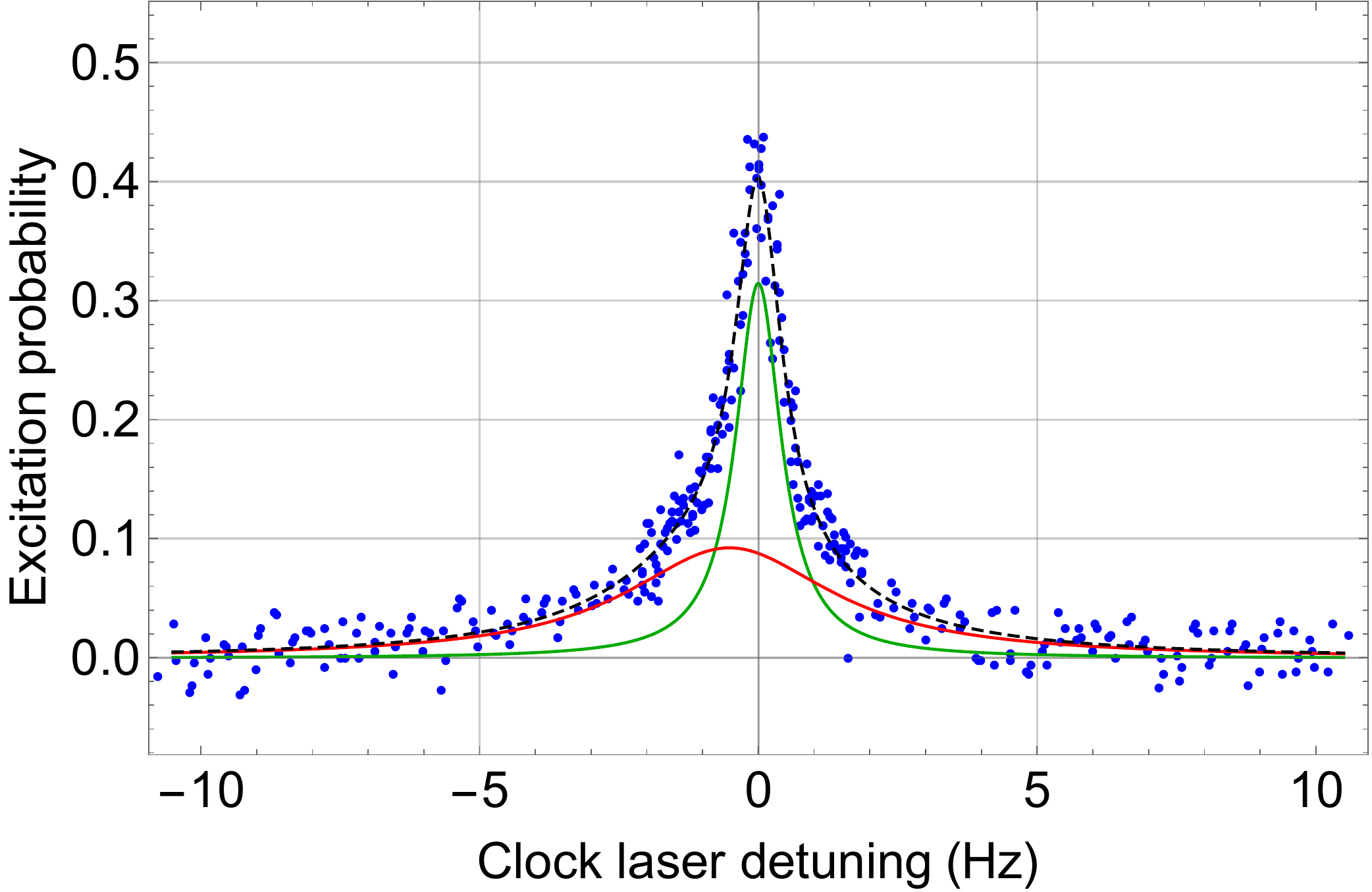} 
\par\end{centering}
}
\begin{center}
\subfloat[]{\begin{centering}
\includegraphics[scale=0.37]{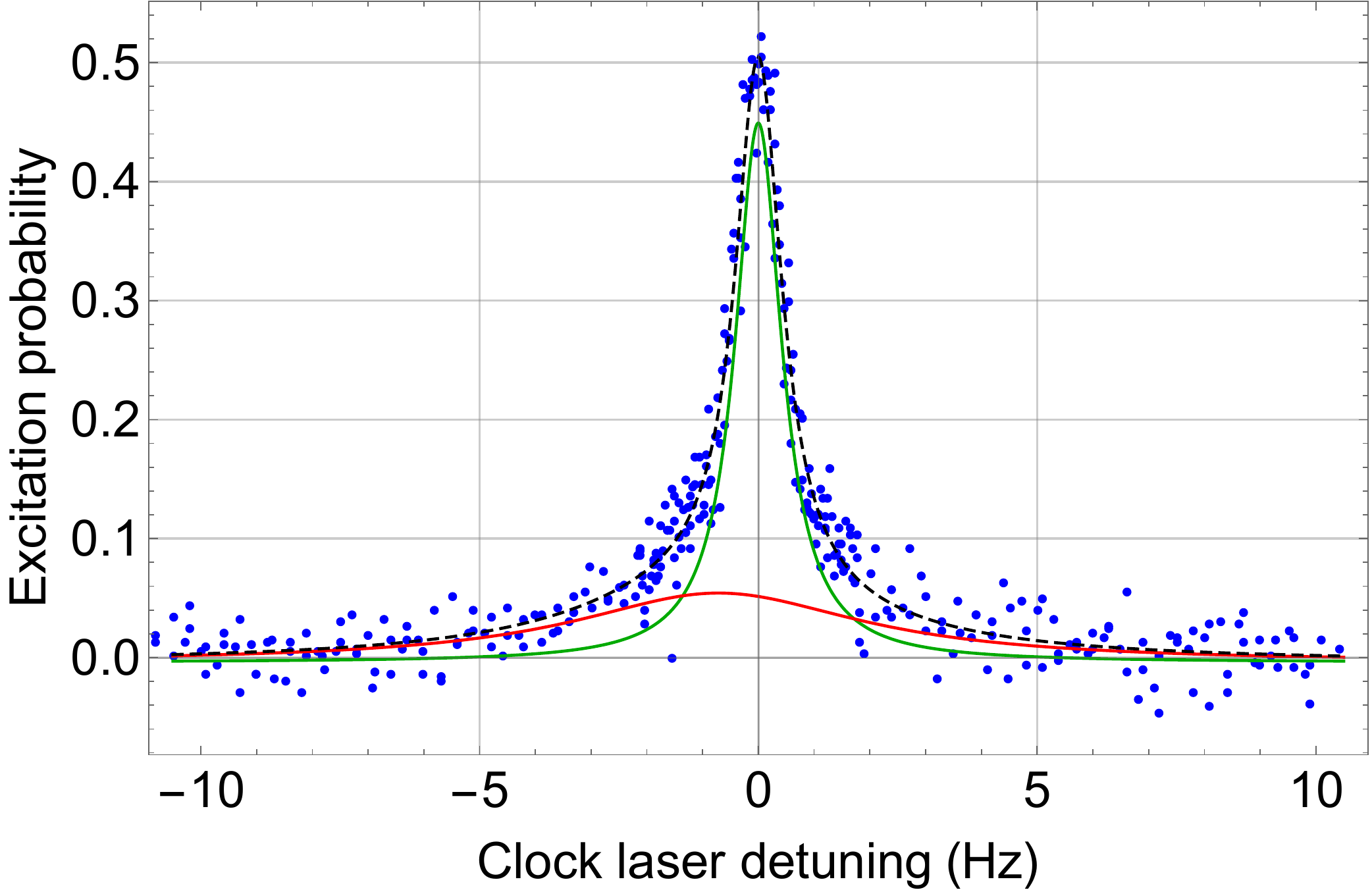} 
\par\end{centering}
}
\par\end{center}%
\end{minipage}
\par\end{centering}
\noindent \caption{Clock transition scans acquired with $T_{\pi}=$1.0~s interrogation
time (a) with detuned photoassociation laser applied before each atom
interrogation, (b) with resonant PA laser. The blue dashed curve is
the fit function including a line from singly occupied sites (green)
and a second line from multiply occupied sites (red). Details are
given in the text.\label{fig:clockTransLineshape}}
\end{figure}

In addition to the ``standard'' determination, we introduce a novel
approach based on a lineshape analysis. Fig.~\ref{fig:clockTransLineshape}
shows two clock transition scans obtained with $T_{\pi}=1.0$~s interrogation
time. Following the works presented in~\citep{Bishof2011,Lemke2011},
which report the observation of atomic interaction sidebands by proper
control of interaction parameters, we model the lineshape as a sum
of two Lorentzian profiles: the main one (green) results from the
atoms in singly-occupied lattice sites, its width is close to the
Fourier limit.
 The second profile (red) results from atoms in multiply occupied
lattice sites. It has a broader width and a negative frequency shift~\citep{Lisdat2009}.
In order to confirm this model, we used photoassociation (PA)~\citep{Zelevinsky2006}:
two atoms in the same lattice site form an excited Sr\textsubscript{2}
molecule by interacting with a photon from the PA beam. The molecule
decays with high probability into two hot atoms, which are lost from
the lattice. In this way, the fraction of multiply occupied sites
is reduced. The PA transition is driven by a radiation detuned by
about $-222$~MHz from the \textsuperscript{1}S\textsubscript{0}-\textsuperscript{3}P\textsubscript{1}
transition~\citep{Takano2017} (689~nm, obtained from the same laser
as employed in the laser cooling) applied for 600~ms before the clock
interrogation, with an intensity of about 1~W/cm\textsuperscript{2}.
During the line scan in Fig.~\ref{fig:clockTransLineshape}(a), the
PA beam frequency is detuned by a few MHz from the PA transition.
This ensures that all disturbances due to the PA beam (such as atom
heating), except for the PA process itself, are equal to the situation
in Fig.~\ref{fig:clockTransLineshape}(b), where the PA laser is
tuned on-resonance. The comparison between the two lineshapes confirms
the model: the contribution of the profile from multiply occupied
sites (red) is reduced when PA is applied. The fact that this contribution
is not fully cancelled could be explained by excited molecules decaying
into the internal ground-state and remaining trapped in sites that
were originally occupied by three or more Sr atoms. These could than
lead to collisional broadening and shifts. Having verified this model,
we can assume that the main profile (green) represents the unperturbed
transition (i.e., atoms not affected by collisions) and evaluate the
collisional shift as the line pulling due to the second profile. Furthermore,
in order to be less sensitive to this line pulling, the main profile
interrogation is done experimentally at two detunings closer to the
center of the main profile than at half height.
\begin{figure}[b]
\begin{centering}
\includegraphics[scale=0.47]{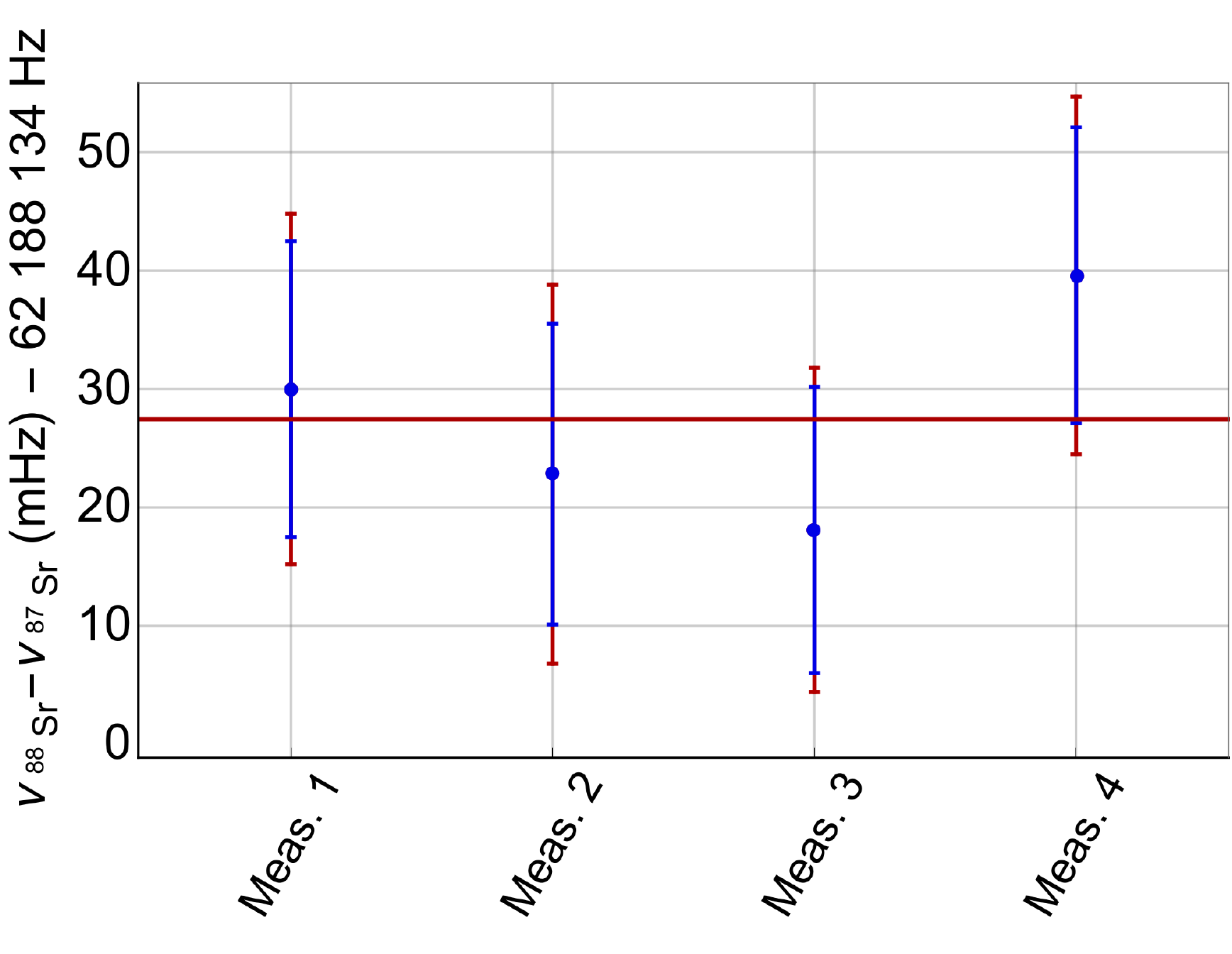} 
\par\end{centering}
\centering{}\caption{Measurement used for the evaluation of the isotope shift value (red
dashed line). The red and blue error bars represent the statistical
and systematic uncertainty. \label{fig:comp-meas-ratio}}
\end{figure}

During the measurement of the isotope shift (see Section~\ref{sec:isotope-shift})
no PA beam was used. However, the atoms were loaded into the lattice
starting from a red MOT with higher temperature and larger size, meaning
lower density, spreading the atoms over more lattice sites. In addition,
the longer interrogation time provides narrower transition lines from
singly occupied sites and reduces the effect of line pulling. The
resulting line pulling during the measurement of the \textsuperscript{88}Sr–\textsuperscript{87}Sr
isotope shift is $\Delta\nu_{\mathrm{LP}}=2.5(15)$ mHz, or $5.7(34)\times10^{-18}$,
consistent with the result obtained with ``standard'' evaluation,
$0.5(22)$~mHz ($1(5)\times10^{-18}$).

\subsection{DC Stark shift }

The DC Stark shift is a potential issue, since the small size of the
vacuum chamber places the non-conductive chamber windows, a potential
location of accumulated electric charges, as close as 7~mm to the
atoms. The shift is measured by applying voltages, in turn, to three
pairs of approximately circular wire electrodes placed externally
to the windows, and measuring the resulting clock transition frequency
shift. From the quadratic fit, the residual shift for the operating
condition of zero applied voltages is $2(2)\times10^{-18}$.

\subsection{Background gas collisions shift}

The background gas collisions shift can be evaluated, as reported
in~\citep{Gibble2013}, from the lattice lifetime and using the coefficients
given in~\citep{Mitroy2010}. The measured lattice lifetime is 5.6~s,
leading to a shift of $-1.3(13)\times10^{-18}$.

\subsection{Tunneling}

We assume the contribution of tunneling to be negligible, since the
lattice beams are vertically oriented and the lattice is deep~\citep{Lemonde2005}.

\section{isotope shift measurement\label{sec:isotope-shift}}

\begin{figure}
\begin{centering}
\includegraphics[scale=0.44]{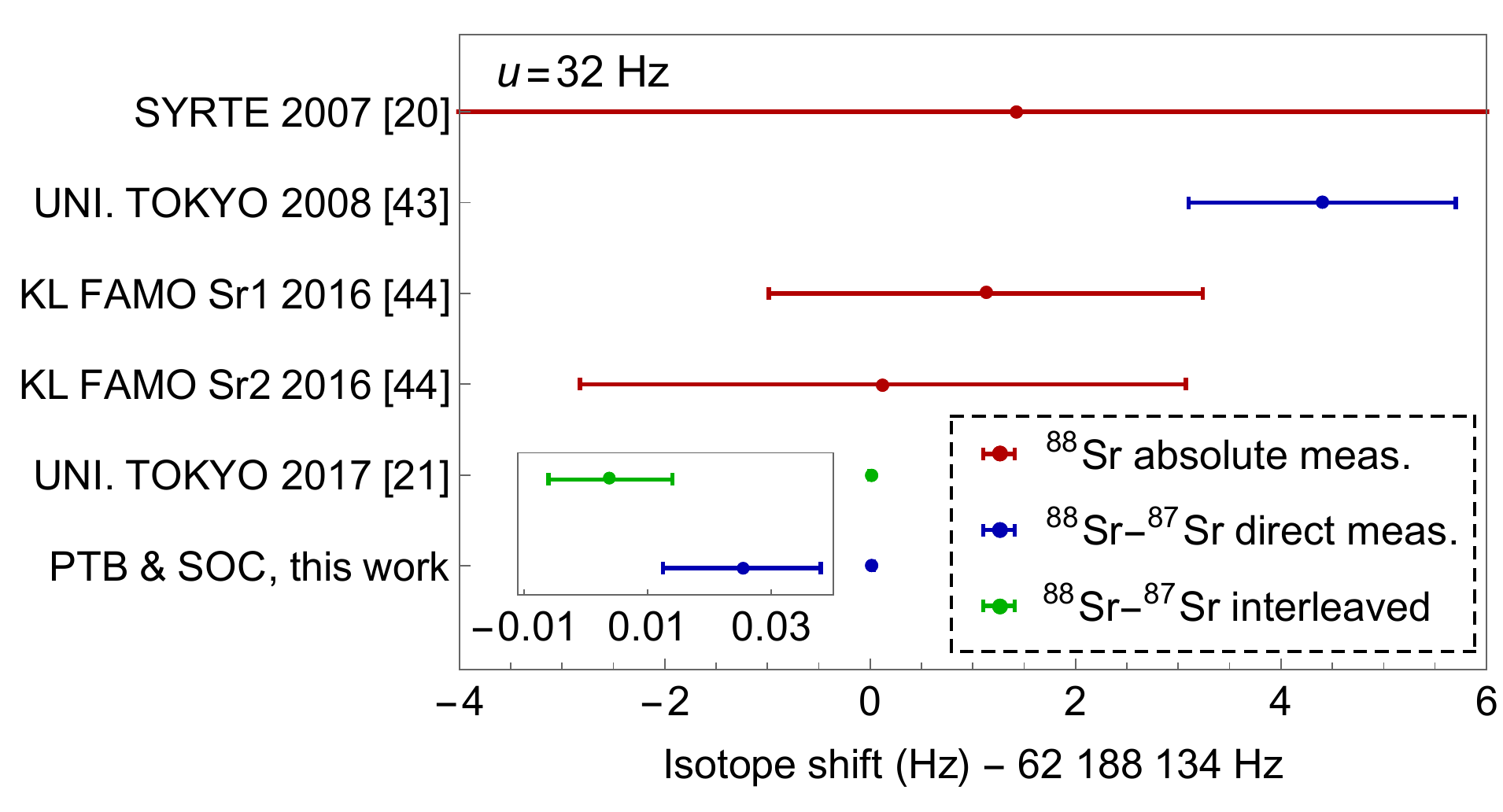} 
\par\end{centering}
\centering{}\caption{Present isotope shift, compared with published values.\label{fig:88Sr87SrIsotopeShiftComparison}}
\end{figure}

For the evaluation of the isotope shift, from the comparison with
the PTB clock, we used the average of four measurements (Fig.~\ref{fig:comp-meas-ratio})
acquired over two days. The lattice shift, probe light shift and the
Zeeman shift measurements were acquired in the same days. The systematic
uncertainty, 12~mHz, arises from the bosonic clock's uncertainty
and the PTB clock's uncertainty, The resulting \textsuperscript{88}Sr–\textsuperscript{87}Sr
isotope shift is 62~188~134.027(12)~Hz, where the uncertainty corresponds
to $3.0\times10^{-17}$. In Fig.~\ref{fig:88Sr87SrIsotopeShiftComparison}
our measurement is compared with previously published values.
 In particular, the difference compared with the value recently reported
in~\citep{Takano2017} is within $2\sigma$ of the combined uncertainty
of both measurements.

\section{Conclusions and outlook}

In conclusion, we have demonstrated a bosonic optical lattice clock
with $3\times10^{-18}$ instability and $2.0\times10^{-17}$ inaccuracy.
This result was obtained by operating the clock with long interrogation
times of 4~s, with suitably low atom density, stabilizing the important
physical parameters of the apparatus with active control, and accurate
evaluation of the shifts. The long interrogation time was only possible
by exploiting the ultra-low instability of a cryogenic reference cavity~\citep{Matei2017}.
As a consequence, we determined the \textsuperscript{88}Sr–\textsuperscript{87}Sr
isotope shift with 12~mHz inaccuracy. With this study, we realize
the long-predicted potential of bosonic lattice clocks, with a factor
of approximately 30 improvement in terms of accuracy and instability
compared with the best values reported so far~\citep{Takano2017}.

We see significant potential for improving this approach further,
by achieving a longer atom lifetime in the lattice, and implementing
for instance higher-dimensional lattices, hyper-Ramsey spectroscopy~\citep{Yudin2010}
and PA. Bosonic clocks may than become competitive with fermionic
ones, mainly for applications where simplicity, reliability, or fundamental
physics are essential. The present technique could also be applied
to other species (Yb, Mg), which have values of $|\Delta\nu_{\mathrm{L}}+\Delta\nu_{\mathrm{B}}|_{{\rm \mathrm{min}}}$
similar to Sr~\citep{Taichenachev2006}.

\begin{acknowledgments}
The authors are very grateful to C. Klempt and I.~Kruse from Leibniz-Universität
Hannover for providing the control software for the FPGA, to D.~Iwaschko
and U.~Rosowski (HHU), and to M.~Misera, A.~Koczwara and A.~Uhde
(PTB) for technical assistance and useful discussion, to T. Legero
and D.G. Matei for making available the cryogenic silicon cavity,
and to E. Benkler for operating the frequency comb. We are indebted
to the members of the SOC2 consortium for their contributions to the
early development of the apparatus described here and for equipment
loan. We thank L. Cacciapuoti (ESA) for constant support. This work
was funded in part by the FP7-MSCA-ITN project No.~607493 ``FACT”,
the H2020-MSCA-RISE- project No.~691156 ``Q-Sense'' and ESA project
No. 4000119716 (``I-SOC''). The PTB team acknowledges funds from
the CRC 1227 DQ-mat within project B02. 
\end{acknowledgments}

\bibliography{ciao}
\bibliographystyle{apsrev4-1}
 {}
\end{document}